\documentclass[11pt,a4paper]{article}
\usepackage{jheppub}
\newcommand\beq{\begin{equation}}
\newcommand\eeq{\end{equation}}
\newcommand\bea{\begin{eqnarray}}
\newcommand\eea{\end{eqnarray}}
\newcommand\bi{\begin{itemize}}
\newcommand\ei{\end{itemize}}
\newcommand\ben{\begin{enumerate}}
\newcommand\een{\end{enumerate}}

\newcommand{\ba}{\begin{array}}
\newcommand{\ea}{\end{array}}
\newcommand{\ee}{\end{equation}}
\newcommand{\beqa}{\begin{eqnarray}}
\newcommand{\eeqa}{\end{eqnarray}}

\def\321{$SU(3)\times SU(2)\times U(1)$}
\def\10{$SO(10)$}
\def\b126{$\overline{126}$}

\newcommand{\dms} {\Delta m^2_{\rm sol}}
\newcommand{\Dma} {\Delta m^2_{\rm atm}}
\newcommand{\mq}[1]{m_{\tilde{#1}_{1,2}}}
\newcommand{\Y}[1]{\mathbf{Y}_{#1}}

\newif\ifboo \boofalse

\def\lsim{\mathrel{\rlap{\lower4pt\hbox{\hskip1pt$\sim$}}
    \raise1pt\hbox{$<$}}}         %less than or approx. symbol
\def\gsim{\mathrel{\rlap{\lower4pt\hbox{\hskip1pt$\sim$}}
    \raise1pt\hbox{$>$}}}         %greater than or approx. symbol

\title{Status of  Supersymmetric Seesaw in  SO(10) models}
\author[1]{L. Calibbi,}
\affiliation[1]{Max-Planck-Institut f\"ur Physik (Werner-Heisenberg-Institut),
M\"unchen 80805, Germany}
\emailAdd{calibbi@mppmu.mpg.de}
\author[2]{ D. Chowdhury,}
\affiliation[2]{Centre for High Energy Physics, Indian Institute of Science, Bangalore 560 012, India}
\emailAdd{debtosh@cts.iisc.ernet.in}
\author[3]{ A. Masiero,}
\affiliation[3]{INFN, Sezione di Padova and Dip. di. Fisica `Galileo Galilei',  Univ. di. Padova, Padova 35131, Italy}
\emailAdd{antonio.masiero@pd.infn.it}
\author[4]{ K. M. Patel}
\affiliation[4]{Physical Research Laboratory, Navarangpura, Ahmedabad 380 009, India}
\emailAdd{kmpatel@prl.res.in}
\author[2]{and S.  K. Vempati}
\emailAdd{vempati@cts.iisc.ernet.in}
\date{\today}                                           % Activate to display a given date or no date

\abstract{
We report on the status of supersymmetric seesaw models in the light of recent experimental results
on $\mu \to e + \gamma$, $\theta_{13}$ and the light Higgs mass at the LHC. SO(10)-like
relations are assumed for neutrino Dirac Yukawa couplings and two cases of mixing, one
large, PMNS-like, and another small, CKM-like, are considered. It is shown that for the large 
mixing case,  only a small range of parameter space with moderate $\tan \beta$ is still allowed. 
This remaining region can be ruled out by an order of magnitude improvement in the current limit on BR($\mu \to e + \gamma$).  
We  also explore a model with non-universal Higgs mass boundary conditions at the high scale.  
It is shown that  the renormalization 
group induced flavor violating slepton mass terms are highly sensitive to the Higgs boundary 
conditions. Depending on the choice of the parameters, they can either lead to 
strong enhancements or cancellations within the flavor violating terms. Such cancellations might
relax the severe constraints imposed by lepton flavor violation compared to mSUGRA. Nevertheless for a large region of parameter space the predicted rates lie within the reach of future experiments once the light Higgs mass constraint is imposed. 
We also update the potential of the ongoing and future experimental searches for lepton flavor violation
in constraining the supersymmetric parameter space. }

\arxivnumber{}
\begin{document}
\maketitle
%\section{}
%\subsection{}

\section{Introduction}
\label{sec1}
Current  times are  unprecedented in terms  of  experimental activity in high energy physics.
There have been several results of very high impact  in the recent times. The following
three are the most relevant for the purpose of our discussion. 

\begin{itemize}
\item Firstly, LHC experiments have reported the discovery of a new boson with mass of about 
125 GeV \cite{cmshiggs,atlashiggs}, compatible with the Standard Model (SM) Higgs boson.
In minimal supersymmetric standard model (MSSM), this would imply a mass of this order and SM-like couplings 
for the lightest CP-even Higgs boson \cite{strumia, Ellis:2012hz,Espinosa:2012im,Carmi:2012in}.
In our analysis, we will take the $2\sigma$ mass range obtained in \cite{strumia}
\begin{equation}
\label{higgs-range}
124.5  ~\text{GeV} \lesssim m_h \lesssim 126.5 ~\text{GeV},
\end{equation}
where $m_h$ stands for the mass of the lightest neutral Higgs. 
In addition, LHC has  also improved the limits on 
the spectrum of low-energy supersymmetry (SUSY) \cite{CMS_2012}. 
%\lorenzo{[add updated refs.]}.

\item The  limit on the lepton flavor violating (LFV) decay $\mu \to e + \gamma$ has improved
by one order of magnitude \cite{Adam:2011ch}. The current limit is 
\begin{equation} 
\text{BR}(\mu \to e + \gamma) < 2.4 \times 10^{-12} ~(90\%~ \textrm{CL}).
\end{equation}
\item  Finally, the so-far unknown neutrino mixing angle, $\theta_{13}$ has been experimentally 
determined \cite{An:2012eh,Ahn:2012nd}. The Daya Bay and RENO experiments have
measured $\theta_{13}$ with very good accuracy: 
\beqa \label{expt-t13}
\sin^2 2 \theta_{13} =& 0.092 \pm 0.016({\rm stat.}) \pm 0.005({\rm syst.})& ~~
\text{Daya Bay \citep{An:2012eh}} \nonumber \\
\sin^2 2 \theta_{13} =& 0.113 \pm 0.013({\rm stat.}) \pm 0.019({\rm syst.})& ~~
\text{RENO \citep{Ahn:2012nd} } \eeqa

\end{itemize} 

The implications of the measurement of the light Higgs mass for various supersymmetric models have been studied in detail by various authors \cite{Hall:2011aa,Heinemeyer:2011aa,Arbey:2011aa,Arbey:2011ab,Draper:2011aa,Carena:2011aa,Christensen:2012ei,
Baer:2011ab,Kadastik:2011aa,Buchmueller:2011ab,Aparicio:2012iw,Ellis:2012aa,Baer:2012uy}. Supersymmetric seesaw models have not been explored so far in the light of these results\footnote{As this work
was finished and being prepared for submission, the following paper Ref.\cite{Hirsch:2012ti} appeared on arXiv.}. Furthermore, flavor violation is expected from SUSY seesaw models where the last two experimental results would play a crucial role in constraining the parameter space. In the present work, we consider the implications of all the three experimental results on a class of SUSY seesaw  models inspired by SO(10) GUTs. 

%In case the seesaw mechanism is implemented in supersymmetric theories, 
%flavor violation is expected to occur in the charged lepton sector too 
%and the last two experimental results would play a crucial role in constraining the 
%parameter space. In the present work, we consider the implications of all the
%three experimental results on a class of SUSY seesaw  models inspired
%by SO(10) GUTs. 

SO(10) GUT Models typically relate up-quark Yukawa matrices with the Dirac neutrino
Yukawa couplings of the Type I seesaw mechanism. In fact, at least one of the 
neutrino Yukawa couplings is expected to be as large as the top Yukawa coupling, as a consequence
of the undelying SO(10) symmetry \cite{Masiero:2002jn}. 
Concerning the mixing structure of the neutrino Yukawa matrix, two extreme cases
can be motivated by simple SO(10) models: large PMNS-like mixing or small CKM-like mixing.
Lepton flavor violation in this class of models has been studied in Refs.~\cite{Masiero:2002jn,Calibbi:2006nq,Calibbi:2011dn}. 
The present work can be considered as an update of these studies.% in the light of the recent experimental results mentioned above. 

The previous works considered mSUGRA/CMSSM-type boundary conditions for the SUSY-breaking soft terms. While universality of the soft terms at the GUT scale is required for the fields belonging to the same {\bf 16} representation of SO(10), the full universal boundary conditions of the CMSSM type are too strong a condition. For example, there is no fundamental reason why the Higgs doublets (that typically are in 10-dimensional representations) and the sfermion soft masses should be degenerate. Thus, strict universality between matter and Higgs fields can be relaxed \cite{Ellis:2002iu}. These models are typically dubbed Non-Universal Higgs Mass models (NUHM). The relaxation of the universality has an important impact on the seesaw generated flavor violating entries of the slepton mass matrix. The magnitude of the RG generated flavor violating entries can either increase or decrease at the leading order due to the interplay between the matter and Higgs mass terms at the GUT scale. Cancellations in the flavor violating entry can indeed relax the LFV constraints on the SUSY parameter space. However, as we are going to see, the Higgs mass and LFV constraints are such that we have similar conclusions as in the mSUGRA, for moderate/large values of $\tan\beta$: an improvement of one order of magnitude in the BR($\mu \to e +\gamma$) bound is sufficient to rule out significant amount of the parameter space. In the present work, we compare and contrast the constraints on SUSY seesaw parameter space with CMSSM/mSUGRA boundary conditions and NUHM boundary conditions. 

We find that in the PMNS case, $\tan\beta$ is restricted between 4 to 20. Lower values of $\tan\beta$ are disfavored by the light Higgs mass constraints whereas higher values are strongly constrained by the present limit on BR($\mu \to e +\gamma$) whose rates can be very large, as a consequence of the sizable observed value of $\theta_{13}$. Furthermore, the one order of magnitude improvement expected in the near future from the MEG experiment would rule out most of the SUSY parameter space accessible at the LHC.  This situation is somewhat relaxed in the NUHM model where we considered the masses of both the Higgs doublets to be the same (the so-called NUHM1 model). We have further updated the future prospects for both the CKM and PMNS cases in mSUGRA/CMSSM  as well as in NUHM1. 

The  paper is organized as follows. In section \ref{sec2}, we recap the SUSY seesaw and discuss the generation of flavor violation in mSUGRA and NUHM1. In section \ref{sec3}, we discuss the details of our numerical analysis. In section \ref{sec4}, we present our results. We conclude with a summary and outlook in section \ref{sec7}. Finally, in appendix \ref{sec6} we describe the proposed future experiments and their expected sensitivity.

\section{Seesaw in mSUGRA and NUHM}
\label{sec2}
The phenomenology of SUSY Type I seesaw mechanism 
with universal boundary conditions (mSUGRA/CMSSM) 
has been studied in many papers (see \cite{Masiero:2004js,reviews1} for a set of recent works). 
Here we review some essential features related to flavor violation for completeness 
and to do a comparison with the case of non-universal Higgs masses. To set the notation, 
the Type I seesaw mechanism is characterized by a superpotential containing the following terms 
\begin{equation}
\label{seesaww}
\mathcal{W} \supset \Y{e} L e^c H_d + \Y{\nu} L \nu^c H_u + {1 \over 2 } M_R \nu^c \nu^c  
\end{equation}
where $L$ ($e^c$) stands for the leptonic doublets (singlets) and $\nu^c$ are the 
right-handed (RH) neutrino superfields (with the generation indices not explictely written). 
$\Y{e}$ and $\Y{\nu}$ are the electron and neutrino (Dirac) Yukawa matrices. 

In models like CMSSM/mSUGRA, the soft terms are assumed to be universal at the Grand Unification
(GUT) scale, $M_{\rm GUT} \sim 2 \times 10^{16}$ GeV. At the weak scale as 
is well known, the soft terms are no longer universal due to the effects of the renormalization group (RG) running. 
The presence of the RH neutrinos of eq.~(\ref{seesaww}) at an intermediate scale contribute to the running and generate flavor  violating entries in the left-handed slepton mass matrix at the weak scale \cite{Borzumati:1986qx}. 
At the leading order these terms can be estimated to be:
\beq
(m^2_{\tilde{L}})_{i\neq j} \equiv \left(\Delta^\ell_{i\neq j}\right)_{LL} \approx - \frac{3m_{0}^{2}  +
A_0^2}{8\pi^{2}} \sum_{k} \left(Y^{*}_{\nu}\right)_{ik} \left(Y_{\nu}\right)_{jk} \log \left(\frac{M_{X}}{M_{R_{k}}}\right),
\label{msugradeltaLL}
\eeq
where  $M_X$ represents the GUT scale and $M_{R_k}$, the scale of the $k^{\text{th}}$ RH neutrino. $m_0$ and $A_0$
stand for the usual universal soft mass and trilinear terms at the high scale.  $\Y{\nu}$, the 
Dirac neutrino Yukawa couplings are free parameters in the Type I seesaw mechanism 
which cannot be completely determined  even after including the complete data on the neutrino mass matrix \cite{Casas:2001sr}.

SO(10) models with their matter representations being 16-dimensional provide
a natural setting for the seesaw mechanisms. Furthermore, they 
provide information about the neutrino Yukawa couplings.  For example, it is known that
as long as we restrict to renormalisable SO(10) models, at least one of the neutrino
Yukawa couplings should be as large as the top Yukawa coupling \cite{Masiero:2002jn}.
Thus with suitable assumptions for the (left-handed) mixing of the Dirac Yukawa Neutrino mass matrix,
one can make predictions for the flavor violation generated at the weak scale from %through the entry given by
eq.~(\ref{msugradeltaLL}). Two extreme scenarios for mixing are typically considered to be present in $\Y{\nu}$  \cite{Hisano:1995cp,Masiero:2002jn,Calibbi:2006nq} :
\begin{align}
	\Y{\nu} &= \Y{u}\qquad\qquad\qquad\textrm{(CKM Case)} \notag \\
	\Y{\nu} &= \Y{u}^{\rm diag}\; {\bf U}_{\rm PMNS}\quad\textrm{(PMNS Case)},
\label{diracnus}
\end{align}
where $\Y{u} = {\bf V}_{\rm CKM} \Y{u}^{\rm diag} {\bf V}_{\rm CKM}^\dagger$. Both these scenarios can
be motivated from concrete models of fermion masses within the  SO(10) 
framework \cite{ Masiero:2002jn,Calibbi:2006nq}.  The flavor violating off-diagonal
entries at the weak scale,  eq.~(\ref{msugradeltaLL}), are then completely determined by assuming $\Y{\nu}$  as in eq.~(\ref{diracnus}). The dominant combinations of Yukawa couplings which enter the radiative generation of $(\Delta^\ell_{i\neq j})_{LL}$ are shown in table (\ref{lfv-deltaLL}). 
Notice that the hierarchical structure of $\Y{\nu}$ dictated by SO(10) determines
that the leading contribution corresponds to third generation particles running in the loop.
Hence, the flavor violating entries $(\Delta_{12})_{LL}$ responsible for $\mu\rightarrow e \gamma$ process and $(\Delta_{13})_{LL}$ responsible for $\tau \to e \gamma$ depend on $U_{e3}\sim \theta_{13}$ in the PMNS case.  The branching ratios of the LFV decays can be roughly estimated to be
\beq \label{brLFV}
\frac{{\rm BR}(l_{i}\rightarrow l_{j}\gamma)}{{\rm BR}(l_{i}\rightarrow l_{j}\nu \bar{\nu})} \approx \frac{\alpha^{3}}{G^{2}_{F}}\,
\frac{\left(\delta_{LL}\right)^{2}_{ij}}{m^{4}_{\rm susy}} \, \tan^{2}\beta
\eeq
where $m_{\rm susy}$ is a typical SUSY mass and 
the flavor violation is as usual parameterize by the following quantity
\beq \label{delta}
\delta {^{f}_{ij}} \equiv \frac{\Delta ^{f}_{ij}}{m^{2} _{\tilde{f}}}.
\eeq
%It can be easily seen from Eqs. (\ref{msugradeltaLL}, \ref{brLFV}) and Eq. (\ref{delta}),  that
%\beq \label{brLFV-t13}
%BR(\mu \rightarrow e \gamma) \propto  |U_{\mu3}|^2 |U_{e3}|^2\, \tan^{2}\beta,
%\eeq
%where $U_{e3}=\sin \theta_{13}$ is (1,3) elements of $U_{PMNS}$ matrix.

\begin{table}[!t]
%\label{yuk-lfv}
\begin{center}
\begin{tabular}{|c|c|c|}
\hline \hline
Generations  & PMNS & CKM  \\ \hline \hline
$\Delta_{12}$ & $Y_t^2 U_{e3} U_{\mu 3}$ & $Y_t^2 V_{td} V_{ts}$   \\ 
%\hline 
$\Delta_{23}$ & $Y_t^2 U_{\mu 3} U_{\tau 3}$ & $Y_t^2 V_{tb} V_{ts}$   \\ 
%\hline 
$\Delta_{31}$ & $Y_t^2 U_{e3} U_{\tau 3}$ & $Y_t^2 V_{td} V_{tb}$  \\ 
\hline 
\hline 
\end{tabular}
 % RG induced off-diagonal entries in the sleptionic mass matrix in CKM and PMNS mixing case. }
\end{center}
 \caption{The dominant combination of neutrino Yukawa couplings which enter eq.~(\ref{msugradeltaLL}) in CKM and PMNS mixing case.}
\label{lfv-deltaLL}
\end{table}
 
Let us now turn our attention to the NUHM1 boundary conditions.\footnote{Lepton flavor violation in NUHM models has been previously studied in \cite{herrero}, where correlations between $\mu \to e + \gamma$ and $\mu \to e$ conversion rates have been discussed.} At the first sight one might expect that such a modification has no significant effect on the LFV amplitudes, except for those due to the modifications in the sparticle spectrum. However, it turns out that this is not the only modification. The flavor mixing structure of the slepton mass matrix can also be strongly affected. The radiatively generated flavor violating entries in eq.~(\ref{msugradeltaLL}) take the following form in NUHM models:
\beq \label{slepton}
\left(\Delta^\ell_{i\neq j}\right)_{LL} \approx - \frac{2m_{0}^{2} + {m^{2}_{H_u}} +
A_0^2}{8\pi^{2}} \sum_{k} \left(Y^{*}_{\nu}\right)_{ik} \left(Y_{\nu}\right)_{jk} \log \left(\frac{M_{X}}{M_{R_{k}}}\right),
\eeq
where $m_{H_u}$ is the soft mass of the up-type Higgs at the high scale. In the present
work, we consider the NUHM1 scenario, i.e.~$m_{H_u} = m_{H_d}$ at the GUT scale. Furthermore, there can
be a relative sign difference between the universal mass terms for the matter fields 
(that we still call $m_0$ with abuse of notation) and the Higgs mass terms at the GUT scale.
This can clearly lead to cancellations (for $m^2_{H_u}\approx -2~ m_0^2$) or enhancements 
(for $m^2_{H_u}\gtrsim~m_0^2$)) in the magnitude of the flavor violating entries at the weak scale 
compared to mSUGRA.

\section{Parameter Range and Phenomenological Constraints}
\label{sec3}
As mentioned earlier, we will consider two sets of boundary conditions for the soft-terms in our numerical analysis. While the mSUGRA is characterized by the standard `four and half' parameters ($m_0$, $M_{1/2}$, $A_0$, $\tan\beta$, sgn($\mu$)), we parametrize the NUHM1 case by $m_{H_u}=m_{H_d} \equiv m_0 - \Delta m_H$. Considering the present and future LHC accessible regions as well as the reach of future flavor physics experiments, we scan the soft parameter space in the following ranges:

%We take the most conserved value of the reactor mixing angle $|U_{e3}|=0.11$ which is the lowest
%value at 3$\sigma$ set by the current observations from RENO experiments \cite{Ahn:2012nd} as well
%as from the global fits of recent neutrino oscillation data \cite{Tortola:2012te}. The remaining two
%angles of PMNS sector are set to their global fit values $\theta_{12}=33.5^o$ and
%$\theta_{23}=45^o$. 

%In order to show the cancellations in LFV processes arise due to non-universal boundary conditions, we study the NUHM scenario characterized by 

\begin{align} \label{prm}
m_0 &\in [0,\ 5]\ \text{TeV} \nonumber \\
\Delta m_H &\in
\begin{cases}
0 & \qquad {\rm for~mSUGRA} \\
[0,\ 5] &  \qquad{\rm for~NUHM1} \\
\end{cases} \nonumber \\
m_{1/2} &\in [0.1,\ 2]\ \text{TeV} \nonumber \\
A_0 &\in [-3 m_0,\ +3m_0] \nonumber \\
\text{sgn}(\mu) &\in \left\lbrace -,+\right\rbrace  
\end{align}
Note that we use the convention in which $m_{H_u}^2 = $ sgn($m_{H_u}$) $\left|m_{H_u}\right|^2$. For this range of the parameter space the first two generations squarks have masses up to $\mq{q} \simeq 7$ TeV and the first two generations sleptons up to 
$\mq{\ell} \simeq 5$ TeV. We include in our scan such spectra beyond the reach of direct SUSY searches at the LHC, 
in order to check the capability of the flavor violating observables in constraining the parameter space.

The numerical analysis is carried out using the SUSEFLAV package \cite{Chowdhury:2011zr}.
It evaluates 2-loop MSSM RGEs with full $3\times3$ flavor mixing effects and also incorporates
one-loop SUSY threshold corrections in all the MSSM parameters. It checks for consistent
Radiative Electroweak Symmetry Breaking (REWSB) by minimizing the one-loop corrected effective
superpotential. The program incorporates the effect of RH neutrinos on MSSM RGEs and
calculates the branching ratios of various LFV processes induced by such RGE effects. %The scale of
%RH neutrinos and their mixing pattern are input parameters which can be set by the users. 
The program also calculates BR($b\rightarrow s \gamma$) in the minimal flavor violation assumptions. We also calculate the BR($B_s\rightarrow \mu^+ \mu^-$) using ISABMM subroutine of ISAJET \cite{Paige:2003mg}. The light Higgs mass is computed using the full two loop corrections of \cite{Degrassi:2001yf,Brignole:2001jy,Dedes:2002dy,slavich}.
%The scans in each case are carried out for three different values of $\tan\beta = 10$ and 40. We evaluate $5\times10^5$ points in each case for fixed $\tan\beta$. 
First, we collect the points which (a) successfully give REWSB, (b) have no any tachyonic sfermions at the weak scale and (c) have the lightest neutralino as Lightest Supersymmetric Particle (LSP). Then we calculate all the LFV observables, BR($b\rightarrow s \gamma$) using the SUSY spectrum evaluated for each point. Finally, we impose the following experimental constraints on the data points we collected.

\beqa \label{expt-bounds}
121.5~ \text{GeV} \leq m_h & \leq\, 129.5~{\rm GeV}     &  %\text{\cite{strumia}}
\nonumber \\
m_{\tilde{\chi}^\pm}~{\rm (lightest~Chargino~mass)} & \geq\, 103.5~{\rm GeV}     &
\text{\cite{Nakamura:2010zzi}} \nonumber \\
{\rm BR}(B_s \rightarrow \mu^+ \mu^-) & < \, 4.5 \times 10^{-9}   &   \text{\cite{Aaij:2012ac}}
\nonumber \\
2.85 \times 10^{-4} \leq {\rm BR}(b \rightarrow s \gamma) & \leq\, 4.24 \times 10^{-4}  & (2\sigma)
\text{\cite{Asner:2010qj}}.  \eeqa
In comparing our predictions for $m_h$ with the experimental range of eq.(\ref{higgs-range}), we take into account 3 GeV of theoretical uncertainty (for a recent discussion see \cite{Arbey:2012dq}). We have not considered the Supersymmetric solution to $(g-2)_\mu$ discrepancy in the present work.
\begin{table}[t]
 \begin{center}
 \begin{tabular}{lcc}
 \hline
 \hline
 LFV Process & Present bound & Near future sensitivity \\
 && of ongoing experiments\\
 \hline
 BR($\mu\rightarrow e \gamma$) & $2.4 \times 10^{-12}$~\cite{Adam:2011ch}
& $10^{-13}$~\cite{Hewett:2012ns}\\
 BR($\mu\rightarrow e e e$) & $1.0 \times
10^{-12}$~\cite{Bellgardt:1987du} & $-$\\
 CR($\mu\rightarrow e$ in Ti) & $4.3 \times 10^{-12}$~\cite{Dohmen:1993mp}
& $-$\\
 BR($\tau\rightarrow e \gamma$) & $3.0 \times 10^{-8}$~\cite{Asner:2010qj}
& $-$\\
 BR($\tau\rightarrow e e e$) & $3.0 \times 10^{-8}$~\cite{Asner:2010qj} &
$-$\\
 BR($\tau\rightarrow \mu \gamma$) & $4.5 \times
10^{-8}$~\cite{Asner:2010qj} & $10^{-8}$~\cite{Hewett:2012ns}\\
 BR($\tau\rightarrow \mu \mu \mu$) & $2.0 \times
10^{-8}$~\cite{Asner:2010qj} & $3\times10^{-9}$~\cite{Hewett:2012ns}\\
 \hline
 \hline
 \end{tabular}
%\lorenzo{Add refs. for future Belle sensitivities?}}
\end{center}
\caption{Present bounds and expected sensitivities on LFV processes.}
\label{lfv-exp}
\end{table}
 
%\begin{table}[t]
% \begin{center}
% \begin{tabular}{lcc}
% \hline
% \hline
% LFV Process & Present bound & Near future sensitivity \\
% && of ongoing experiments\\
% \hline
% BR($\mu\rightarrow e \gamma$) & $2.4 \times 10^{-12}$~\cite{Adam:2011ch} & $10^{-13}$\\
% BR($\mu\rightarrow e e e$) & $1.0 \times 10^{-12}$~\cite{Bellgardt:1987du} & $-$\\
% CR($\mu\rightarrow e$ in Ti) & $4.3 \times 10^{-12}$~\cite{Dohmen:1993mp} & $-$\\
% BR($\tau\rightarrow e \gamma$) & $3.0 \times 10^{-8}$~\cite{Asner:2010qj} & $-$\\
% BR($\tau\rightarrow e e e$) & $3.0 \times 10^{-8}$~\cite{Asner:2010qj} & $-$\\
% BR($\tau\rightarrow \mu \gamma$) & $4.5 \times 10^{-8}$~\cite{Asner:2010qj} & $10^{-8}$\\
% BR($\tau\rightarrow \mu \mu \mu$) & $2.0 \times 10^{-8}$~\cite{Asner:2010qj} & $3\times10^{-9}$\\
% \hline
% \hline
% \end{tabular}
% \label{lfv-exp}
% \caption{Present bounds and expected sensitivities on LFV processes.} %\lorenzo{Add refs. for future Belle sensitivities?}}
% \end{center}
% \end{table}

In our study, we assume normal hierarchy in the light neutrino mass spectrum and set
\begin{align} \label{numasses}
m_{\nu_1} = 0.001~\text{eV},~~m_{\nu_2} &= \sqrt{\dms + m_{\nu_1}^2}~\text{and}~m_{\nu_3} =
\sqrt{\Dma + m_{\nu_1}^2} 
\end{align}
where $\dms$ and $\Dma$ are the solar and atmospheric squared mass differences respectively and we use
the central values obtained from recent global fits on neutrino data \cite{Tortola:2012te}: 
\begin{align}
 \dms = 7.62 &\times 10^{-5}\, {\rm eV}^2\ {\rm and}\  \Dma = 2.53 \times 10^{-3}\, {\rm eV}^2
\end{align}
Regarding the mixing angles, we take the most conservative value for the reactor mixing angle and set $|U_{e3}|=0.11$, that 
corresponds to the lower limit of the 3$\sigma$ range given by RENO \cite{Ahn:2012nd} as well as by the global fits \cite{Tortola:2012te}. The remaining two angles of the PMNS matrix are set to their global fit values $\theta_{12}=33.5^\circ$ and $\theta_{23}=45^\circ$. The masses of heavy neutrinos which we use in our analysis are:
\beq \label{MRs}
M_{R_1} = 10^6~\text{GeV},~~M_{R_2} = 10^9~\text{GeV},~~\text{and}~~M_{R_3} =
10^{14}~\text{GeV}. \eeq 
%
%as obtained from the seesaw formula imposing the conditions in eq.~(\ref{diracnus}).
%\lorenzo{[Check this statement, this actually true for the PMNS case only...]} 

The recent results from the MEG collaboration \cite{Adam:2011ch} have improved
the existing bound on BR($\mu\rightarrow e \gamma$) by one order of magnitude. The present limits on different LFV
observables are summarized in table \ref{lfv-exp}.
In the following subsections, we discuss in detail the results of numerical analysis
carried out in the PMNS and CKM cases with mSUGRA and NUHM1 boundary conditions and present 
a quantitative comparisons between them. %differentscenarios.\\

\section{LFV in SUSY SO(10)}
\label{sec4}

Let us start considering the PMNS case, where there is a direct link between LFV processes and
neutrino parameters. Besides the CP violating phases, $\theta_{13}$ was the only unknown parameter in the leptonic mixing sector till some time ago. There have been various theoretical models based on the idea that $\theta_{13}$ could
be very small, close to zero. In such a case, the resulting $\mu \to e + \gamma$ and
$\tau \to e + \gamma$ rates could have been suppressed \cite{Antusch:2006vw,Calibbi:2006ne}.  
Recent experiments prove the contrary. 
Both the experiments measuring $\theta_{13}$ are in good agreement with each other and indicate
a sizeable value of $\theta_{13}$. The recent global fit analysis \cite{Tortola:2012te} also leads to similar
value of $\theta_{13}$. All these results indicate that the smallest value of $|U_{e3}|$ is 0.11 at
$3\sigma$. The resulting rates of BR$\left(\mu \to e \gamma\right)$ are significantly enhanced for such a value of $\left|U_{e3}\right|$. As we are going to see, the updated MEG limit together with a large $\theta_{13}$ puts significant 
constraints on mSUGRA for moderate as well as large value of $\tan\beta$.\footnote{The interplay between large $\theta_{13}$ and BR$\left(\mu \to e \gamma\right)$ in the context of discrete flavor groups have been recently discussed in \cite{Altarelli:2012bn}. }
\begin{figure}[t]
\centering
 \includegraphics[width=0.45\textwidth]{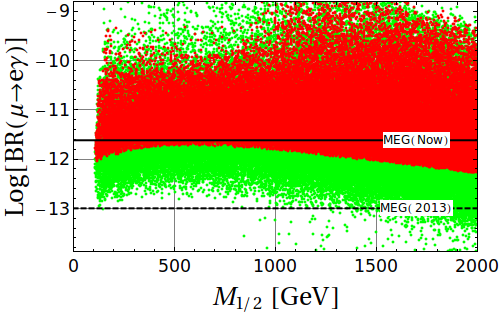}
\hspace{0.3cm}
 \includegraphics[width=0.45\textwidth]{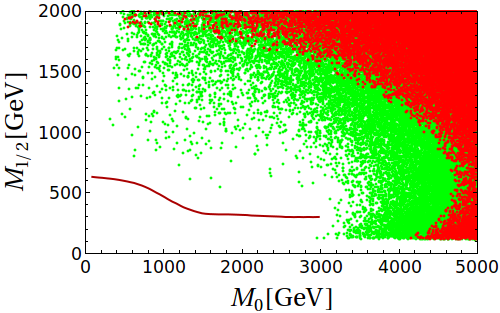}
\caption{The figure in the left panel shows the BR($\mu\rightarrow e \gamma$) obtained by
scanning the mSUGRA (in red color) and NUHM (in green color) parameters in the ranges given in eq.~(\ref{prm}) and for fixed $\tan\beta=10$ and $U_{e3}=0.11$ (the lowest value allowed at $3\sigma$ by
recent RENO observation) and satisfy all the constraints in eq.~(\ref{expt-bounds}). Different horizontal lines correspond to present and future bounds on BR($\mu\rightarrow e \gamma$). The figure in the right
panel shows the allowed space in the $m_0-m_{1/2}$ plane which satisfy the current MEG bound and eq.~(\ref{expt-bounds}). The region below the red line is excluded by the direct searches for SUSY at the LHC \citep{CMS_2012}.}
\label{figtb10}
\end{figure}
\begin{figure}[t]
\centering
 \includegraphics[width=0.45\textwidth]{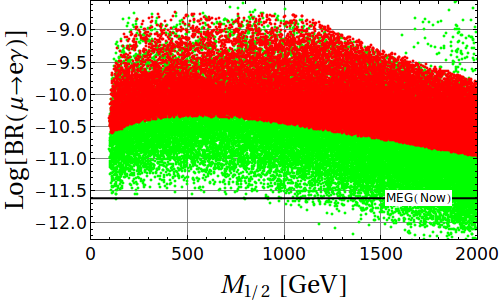}
\hspace{0.3cm}
 \includegraphics[width=0.45\textwidth]{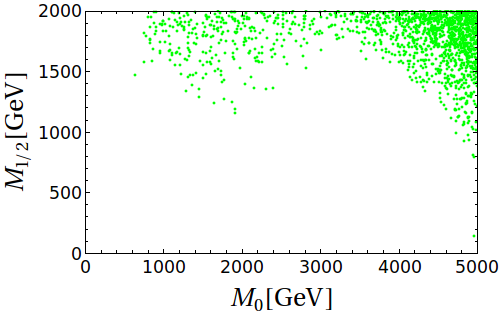}
\caption{The same as figure \ref{figtb10} for $\tan\beta=40$.}
\label{figtb40}
\end{figure}

In figures \ref{figtb10} and \ref{figtb40} we present the constraints from BR$\left(\mu \to e \gamma\right)$ on mSUGRA and NUHM1 parameter space for $\tan\beta=10$ and 40 respectively. 
As can be seen, while only small part of the paramater space survives for $\tan\beta =10$ in mSUGRA, it is completely ruled out for $\tan\beta = 40$. The allowed regions for low $\tan\beta$ require very heavy spectra, i.e. $m_0 \gtrsim 4$ TeV for small 
$M_{1/2}$ or $M_{1/2} \gtrsim 2$ TeV for small $m_0$. What is surprising is that the constraint on the NUHM1 parameter space 
is not as weak as one might expect form eq.~(\ref{slepton}). As we can see from the figures even in the presence of partial cancellations, most of the NUHM1 parameter space is going to be explored by MEG. If one removes the light Higgs mass constraint, points with stronger cancellations would be allowed, even with $\mu\to e \gamma$ rates below the MEG sensitivity. 
Thus points compatible with the Higgs mass bound, eq.~(\ref{expt-bounds}), do not allow strong cancellations in the flavor violating entry in eq.~(\ref{slepton}). For the large $\tan\beta$ case, the $\mu\to e \gamma$ constraint is so strong that
only few points with $M_{1/2} \gtrsim 800$ GeV are allowed. 
In the section \ref{secHiggs}, we will discuss in more detail about the impact of the constraint on $m_h$ in 
mSUGRA and NUHM1.
\begin{figure}[ht]
 \includegraphics[width=0.45\textwidth]{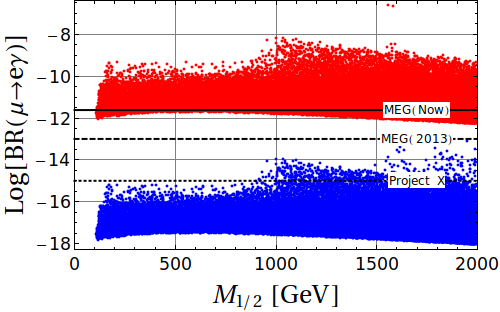}
\hspace{0.3cm}
 \includegraphics[width=0.45\textwidth]{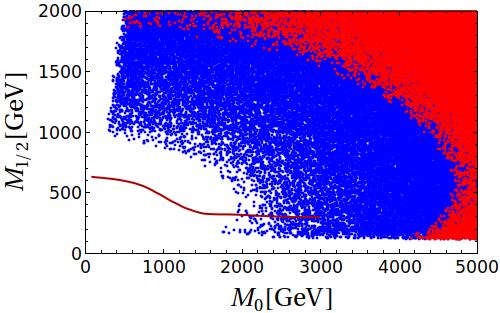}
\caption{The figure in the left panel shows the BR($\mu\rightarrow e \gamma$) obtained by scanning
the mSUGRA parameters in the ranges given in eq.~(\ref{prm}) and for fixed $\tan\beta=10$ and
$U_{e3}=0.11$ (the lowest value allowed at $3\sigma$ by recent RENO observation). The red
(blue) colored points correspond to PMNS (CKM) case. Different horizontal lines correspond to
present and future bounds on BR($\mu\rightarrow e \gamma$). The figure in the right panel shows the allowed space in the $m_0-m_{1/2}$ plane which satisfy the current MEG bound.  The region below the red line is excluded by the current LHC searches \cite{CMS_2012}. Both the plots satisfy all the constraints in eq.~(\ref{expt-bounds}).}
\label{figtb10-ckm}
\end{figure}

% \begin{figure}[h!]
%  \includegraphics[width=8cm]{tb40-ckm-fig1.png}
% \hspace{0.2cm}
%  \includegraphics[width=8cm]{tb40-ckm-fig2.png}
% \caption{The figure in the left panel shows the BR($\mu\rightarrow e \gamma$) obtained by scanning
% the mSUGRA parameters in the ranges given in eq.~(\ref{prm}) and for fixed $\tan\beta=40$ and
% $U_{e3}=0.11$ (the lowest value allowed at $3\sigma$ by recent RENO observation). The red
% (blue) colored points correspond to PMNS (CKM) case. Different horizontal lines correspond to
% present and future bounds on BR($\mu\rightarrow e \gamma$). The figure in the right panel shows the
% allowed space in the $m_0-m_{1/2}$ plane which satisfy the current MEG bound. Both the plots satisfy all the constraints in eq.~(\ref{expt-bounds}).}
% \label{figtb40-ckm}
% \end{figure}

In the context of the updated MEG limit on BR($\mu\rightarrow e \gamma$), it is now worthwhile to see
what is the situation with the small mixing CKM case. Here we compare the CKM case and the PMNS case with mSUGRA boundary
conditions. As above, red points correspond to the PMNS case while we use the blue color for CKM case. The CKM case has highly suppressed branching fractions due to the smallness of CKM angles (see table (\ref{lfv-deltaLL})) as has been detailed in \cite{Calibbi:2006nq}. Though there has been no strong improvements in the experimental sensitivity compared to the analyses of \cite{Calibbi:2006nq}, we update the result with the light Higgs mass constraint.
%For $\tan\beta=10$ (figure \ref{figtb10-ckm}) and 40 (figure \ref{figtb40-ckm}), some part of CKM region can be probed by the proposed Project-X experiment for $\mu\rightarrow e \gamma$. The CKM case for $\tan\beta=10$ allows some solutions for low $m_0$ and $m_{1/2}$ (not shown in figure). Such a region with $m_{1/2}<350$ GeV and $m_0<1$ TeV is ruled out by the current LHC searches \lorenzo{[Are these numbers still valid?]}. The current LHC experiment, operating at $\sqrt{s}=8$ TeV, can exclude some more region in $m_0-m_{1/2}$ plane. For $\tan\beta=40$, the MEG limit requires either $m_0>1.2$ TeV for $m_{1/2}>1.5$ TeV.
%
In figure \ref{figtb10-ckm} we show the results for $\tan\beta=10$. As we can see, some part of the parameter space of the CKM 
case can be probed by the proposed Project-X experiment\footnote{In appendix \ref{sec6} we present a brief summary of all the future experimental facilities related to the flavor violating observables discussed in the text.} for $\mu\rightarrow e \gamma$. At present the main constraint to 
this scenario is simply provided by the $m_h$ range of eq.~(\ref{higgs-range}), that excludes the regions with lighter SUSY spectra: $m_0 \lesssim 2$ TeV for small $M_{1/2}$, $M_{1/2} \lesssim 1$ TeV for small $m_0$, as we can see from the right panel of the figure.
We can also notice that the LHC limits on the mSUGRA parameter space has already started to constrain regions of the parameter space
otherwise allowed by the bounds in eq.~(\ref{expt-bounds}). 

\begin{figure}[ht]
\begin{center}
\includegraphics[width=0.45\textwidth]{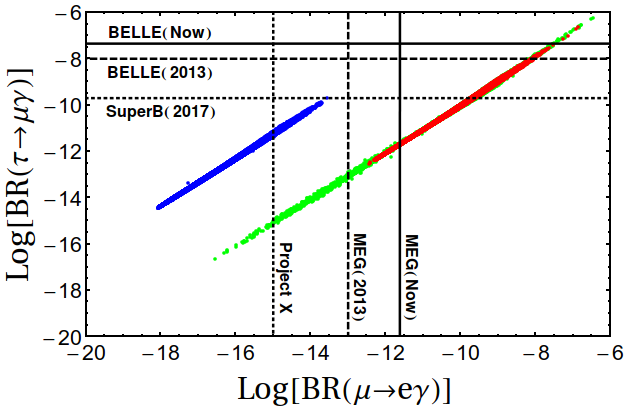}
\hspace{0.3cm}
\includegraphics[width=0.45\textwidth]{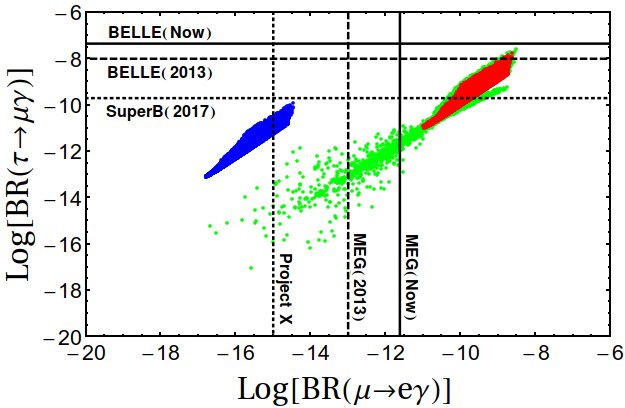}
\end{center}
\caption{BR($\tau \rightarrow \mu \gamma$) versus BR$(\mu \to e \gamma)$ for the PMNS case in mSUGRA (red) and NUHM (green), and for
the CKM case (blue) for $\tan\beta$=10 (left), for $\tan\beta$=40 (right). The different horizontal and
vertical lines correspond to present and future limits on the respective observables.}
\label{fig-tmg}
\end{figure}

\begin{figure}[ht]
\begin{center}
\includegraphics[width=0.45\textwidth]{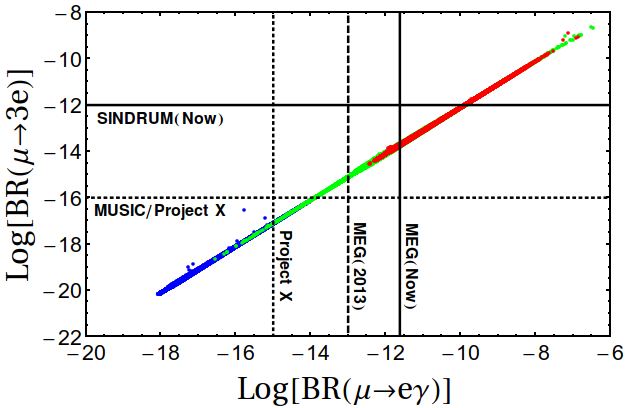}
\hspace{0.3cm}
\includegraphics[width=0.45\textwidth]{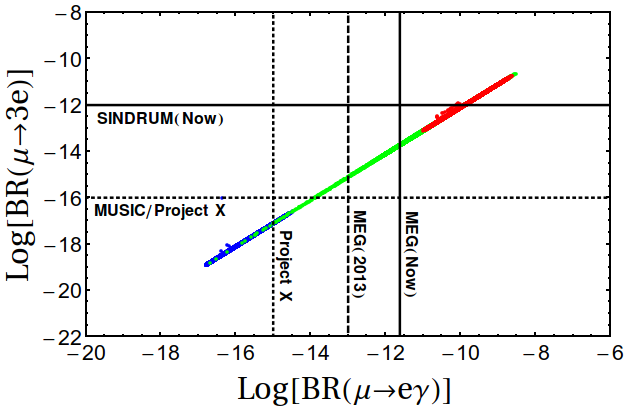}
\end{center}
\caption{BR($\mu \rightarrow eee$) versus BR$(\mu \to e \gamma)$ for the PMNS case in mSUGRA (red) and NUHM (green), and for
the CKM case (blue) for $\tan\beta$=10 (left), for $\tan\beta$=40 (right). The different horizontal and
vertical lines correspond to present and future limits on the respective observables.}
\label{fig-mu3e}
\end{figure}

\begin{figure}[ht]
\begin{center}
\includegraphics[width=0.45\textwidth]{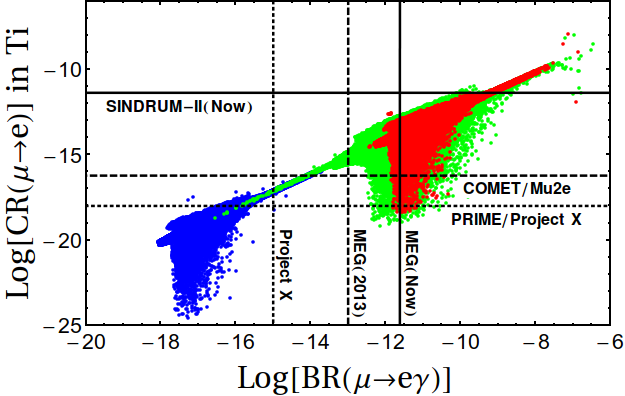}
\hspace{0.3cm}
\includegraphics[width=0.45\textwidth]{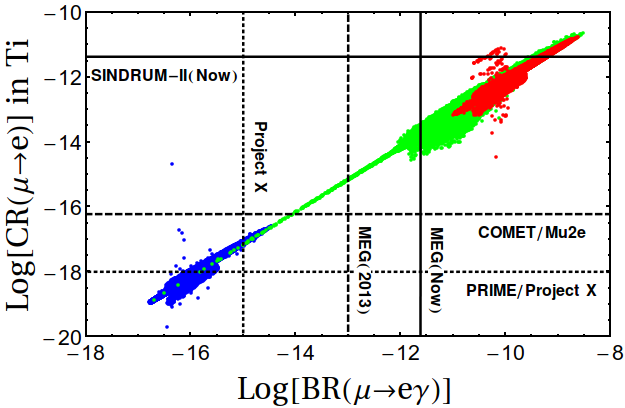}
\end{center}
\caption{CR($\mu \rightarrow e$ in Ti) versus BR$(\mu \to e \gamma)$ for the PMNS case in mSUGRA (red) and NUHM (green), and for
the CKM case (blue) for $\tan\beta$=10 (left), for $\tan\beta$=40 (right). The different horizontal and
vertical lines correspond to present and future limits on the respective observables.}
\label{fig-mutoe}
\end{figure}

Let us now turn our attention to other observables like $\mu \to eee$, $\mu \to e$ conversion in nuclei and $\tau \to \mu \gamma$, 
which is independent of $\theta_{13}$. In figures \ref{fig-tmg}, \ref{fig-mu3e} and \ref{fig-mutoe}, we show the
predicted rates for $\tau \to \mu \gamma$, $\mu \to eee$ and $\mu \to e$ conversion in the Titanium nucleus 
versus the BR$(\mu \to e \gamma)$ (that is at present the most constraining LFV observable), 
for the PMNS case in mSUGRA (red points) and in NUHM1 (green points) as well as for the CKM case (blue points).

As can be seen from figure \ref{fig-tmg}, in the PMNS case, the present MEG limit on BR$\left(\mu \to e \gamma\right)$ implies 
BR$(\tau\rightarrow \mu \gamma)\lesssim 10^{-12}$, beyond the reach of the proposed experiments. This is a direct consequence of the large value of $\theta_{13}$ measured by Daya Bay and RENO. In fact, from eq.(\ref{brLFV},\ref{delta}) and table \ref{lfv-deltaLL}, we have:
\begin{equation}
 \frac{{\rm BR}(\tau\rightarrow \mu \gamma)}{{\rm BR}(\mu\rightarrow e \gamma)} \approx 
\frac{|U_{\tau 3} U_{\mu 3}|^2}{|U_{\mu 3} U_{e3}|^2}\times {\rm BR}(\tau\rightarrow \mu \nu \bar{\nu}) \approx \mathcal{O}(1).
\end{equation}
In the CKM case (blue points), the small mixing angle and the $m_h$ bound are such that 
BR$(\tau\rightarrow \mu \gamma)\lesssim 10^{-10}$.
Thus, the scenarios discussed here allow possible signals of LFV in $\mu-e$ transition only and 
evidence for ${\rm BR}(\tau\rightarrow \mu \gamma)$ at future experiments would strongly disfavor them.

Let us now consider the decay $\mu \rightarrow e e e$. It is known (see e.g. \cite{Hisano:2009ae}) that in SUSY (with conserved R-parity) the
dominant contribution to this process arises from the same dipole operator responsible for $\mu \to e \gamma$,
hence the correlation of the two processes is striking:
\begin{equation}
 {\rm BR}(\mu\to eee) \sim \alpha_{\rm em} \times {\rm BR}(\mu\to e\gamma).
\end{equation}
Such prediction is consistent with our results shown in figure \ref{fig-mu3e} for $\tan\beta=10$ and 40.
The present bound on $\mu \rightarrow e e e$ comes from the SINDRUM experiment at PSI. 
At present, MEG sets a stronger bound, as expected.
However the future sensitivity reach of MUSIC and Project-X experiments will be able to go
beyond the reach of MEG, testing most of the NUHM parameter space of our scan.

Let us now discuss $\mu \to e $ conversion in Nuclei, that will represent one of the most important probes 
of LFV in the future. Project X and J-PARC are envisaging facilities where $\mu$ conversion on various
Nuclei can be studied (for a review, please see \cite{Kuno:1999jp}). In the present work, we 
have computed the $\mu \to e$ conversion rate in Titanium. The conversion rate is as usual normalized by
the capture rate of the muon by the nucleus. 

In figure \ref{fig-mutoe}, we present our results for CR($\mu \to e~{\rm~in~Ti}$)
with respect to  $\mu \to e +\gamma$. We find that there is a significant
spread in the parameter space. This spread is due to the existence of cancellations
between the penguin contributions at low $M_{1/2}$ and in low tan$\beta$ regions, which
has been noted earlier in the literature \cite{Hisano:1995cp}. Still, we can see
that the future experiments will be able to test most of the PMNS parameter space 
and will start to constrain the small-mixing scenario (CKM case). 

The above plots have been obtained for the Titanium Nuclei. As in the case of $\mu\to eee$, it has been noted that the dominant contributions are from the dipole operators. In such a limit, where only the dipole operators contribute, one can easily estimate the conversion rate for the new nuclei, X, by knowing its effective charge, $Z_{eff}$, the form factor $F(q)$ and the atomic number $Z$ and multiplying the conversion rates presented in the above plots by the ratio:
\begin{equation}
R = \frac{\left[Z_{eff}^4 \left|F(q)\right|^2 Z \right]_{\rm X}} {\left[Z_{eff}^4 \left|F(q)\right|^2 Z \right]_{\rm Ti}}
\end{equation}
The Form factors and $Z_{eff}$ for various Nuclei can be found in \cite{Kitano:2002mt}.

\subsection{Combined effect of the Higgs mass and BR$\left(\mu\rightarrow e \gamma\right)$ bounds}
\label{secHiggs}
\begin{figure}[ht]

\includegraphics[width=0.45\textwidth]{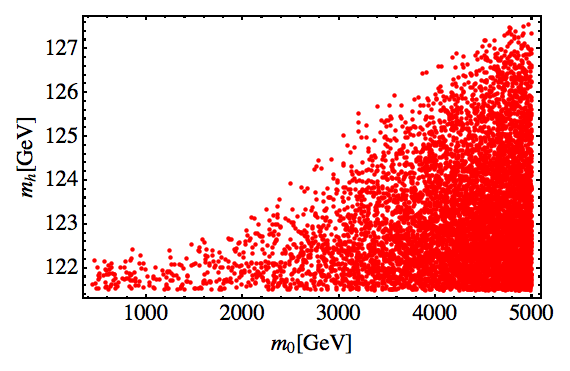}
\hspace{0.3cm}
 \includegraphics[width=0.45\textwidth]{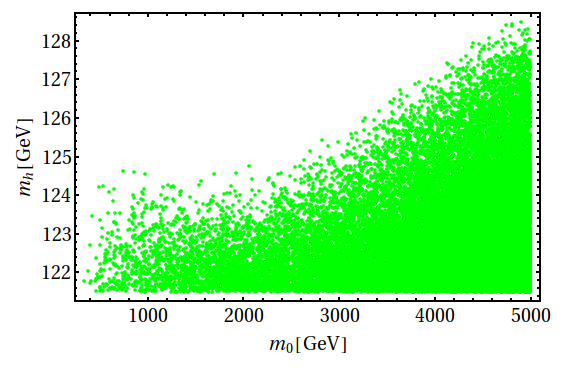}

 \includegraphics[width=0.45\textwidth]{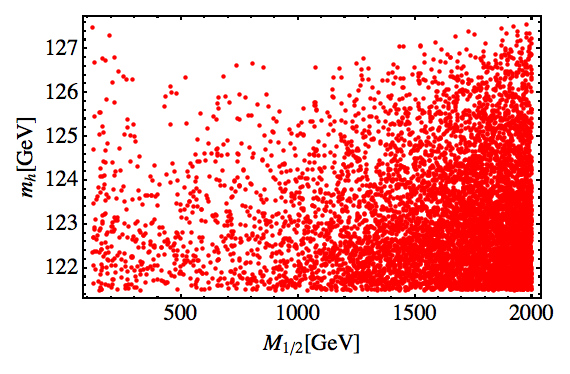}
\hspace{0.3cm}
 \includegraphics[width=0.45\textwidth]{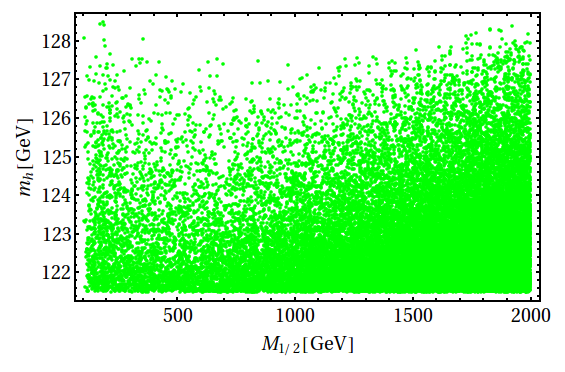}

\vspace{0.3cm}
 \includegraphics[width=0.45\textwidth]{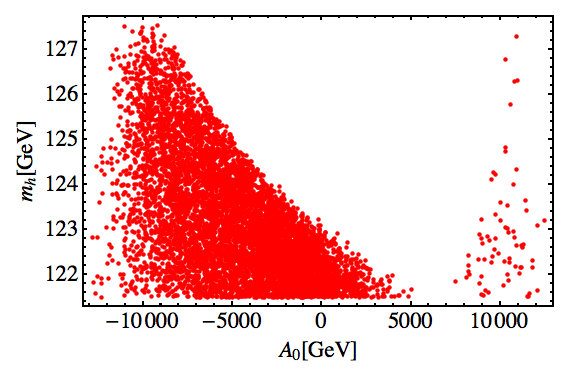}
\hspace{0.3cm}
 \includegraphics[width=0.45\textwidth]{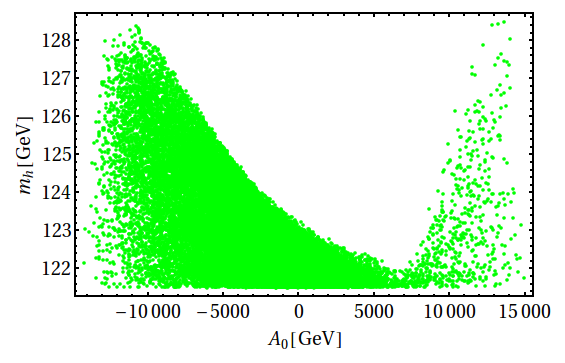}

\vspace*{0.3cm}
 \includegraphics[width=0.45\textwidth]{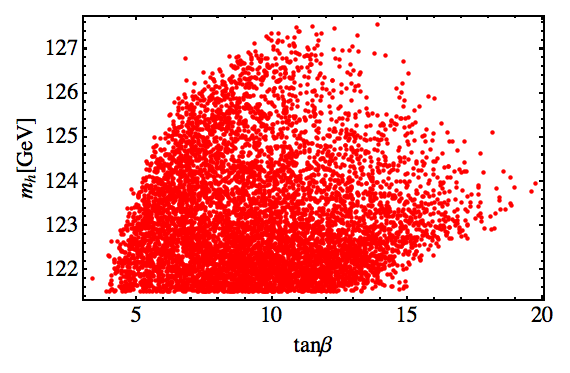}
\hspace{0.3cm}
 \includegraphics[width=0.45\textwidth]{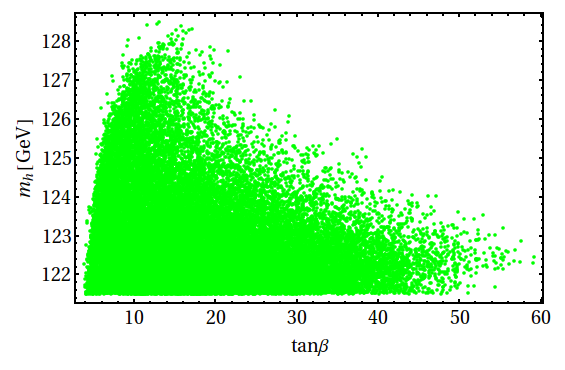}
\caption{Here we present combined regions of parameter space allowed by BR$\left(\mu\rightarrow e \gamma\right)$ and the light Higgs mass ($m_h$), eq.~(\ref{expt-bounds}), on the PMNS case in mSUGRA and NUHM1. } %implications of the light Higgs mass on the 

\label{fig-mh}
\end{figure}

Given the strong constraints from both BR$\left(\mu\rightarrow e \gamma\right)$ and the light Higgs mass one would wonder how much of the total parameter space from eq.~(\ref{prm}) survives in the PMNS case. In figure \ref{fig-mh} we plot the lightest Higgs mass as a function of $m_0$, $M_{1/2}$, $A_0$ and $\tan\beta$ in the three left (right) panels for the mSUGRA (NUHM1) case.
In particular, the left panel of the third row shows the asymmetric regions in $A_0$ required by the Higgs mass range.\footnote{We remind that the gluino-driven radiative effects gives a negative contribution to the stop A-term $A_t$, 
so that, if $A_0 > 0$ only very large values could still provide a sizeable $|A_t|$ at low energy.} Flavor constraints instead are approximately symmetric in $A_0$. 
The last row in the left panel shows the constraint on $\tan\beta$. As we can see in the mSUGRA case both low $\tan\beta$ $(\lesssim 5)$ and high $\tan\beta$ $(\gtrsim 20)$ are ruled out. The constraints in the low $\tan\beta$ are due to $m_h$ whereas those at high $\tan\beta$ are from BR$\left(\mu\rightarrow e \gamma\right)$.

%\lorenzo{[I do not understand much the following argument]}
In mSUGRA/CMSSM models, a light Higgs mass around $\sim$ 125 GeV requires either very heavy stops ($\tilde{t}_{1,2}) \sim 4$ TeV or a large stop mixing triggered by a large low-energy value of the stop $A$-term, $|A_t|$. In either of these cases, one can easily convince oneself, using eq.(\ref{msugradeltaLL}) 
 that the  flavor violating  parameter $\left( \delta^\ell_{i \neq j} \right)_{LL}$ is not suppressed.  Thus a light Higgs mass $\sim 125$ GeV does not necessarily
  mean a suppressed flavor violating  entry in spite of the largeness of stops ($\tilde{t}_{1,2}$) or $A$-terms required. In fact flavor violation constraints are still very strong.%\begin{align}
%\label{brlfv-sim}
%&{\rm BR}(l_{i}\rightarrow l_{j}\gamma) \approx \left|C_{ij}\right|^2 \ \frac{\left(3 m_0^2 + A_0^2 \right)^2}{m_{\rm SUSY}^4} \\
%\intertext{where}
%C_{ij} &= - \frac{1}{8\pi^{2}}\  \frac{\alpha^{3}}{G^{2}_{F}}\  \tan^{2}\beta\  \sum_{k} Y^{\nu*}_{ik} Y^{\nu}_{jk}\ \log \left(\frac{M_{X}}{M_{R_{k}}}\right)
%\end{align} 

%From the above eq.~(\ref{brlfv-sim}) we can easily see that the branching ratio goes to a limiting value of $\mathcal{O}(1)$ in either of the cases where the light Higgs mass $\mathcal{O}(125)$ GeV is realized. Thus leading to the conclusion that Higgs mass $\sim$ 125 GeV leads to large branching ratio. 

In NUHM1 case this correlation is somewhat lost due to partial cancellation in the flavor violating entry\footnote{In NUHM1 case the cancellations are constrained by the parameter choice of eq.(\ref{prm}).}. All values of $\tan\beta$ are now allowed and $A_0$ is slightly more symmetric compared to the mSUGRA case. The surprising thing is that imposing the light Higgs mass constraint restricts the parameter space to be within the reach of MUSIC and Project-X proposals.

\section{Summary and  Outlook}
\label{sec7}
The discovery of a Higgs-like boson at the LHC with a mass close to 125 GeV 
is one of the most significant achievements in particle physics of all times. In the present work, we 
assumed that the particle seen at the LHC is the lightest neutral Higgs scalar of the MSSM. 
We then studied the implications of the observed mass range on SUSY seesaw models
along with recent improvements in $\text{BR}(\mu \to e + \gamma)$ and measurements 
of $\theta_{13}$.  We looked at Type I seesaw model assuming SO(10) relations between
the top Yukawa and the heaviest RH neutrino Dirac mass. We assuming two 
extreme cases of mixing to be present in the Dirac Yukawa matrix (a) small CKM like
mixing and (b) PMNS like mixing.  

We find that there is a strong complementarity in the PMNS case between the light
higgs mass constraint and  BR($\mu \to e + \gamma$). The lower $\tan\beta$ regions
are strongly constrained by the recent measurement of the light Higgs mass and
high $\tan\beta$ regions by present limit of MEG. What is surprising is that $\tan\beta > 20$ 
is already ruled out.  

The NUHM1 boundary conditions is one more interesting framework where  the interplay
between the light Higgs mass constraint and the LFV constraints comes to play.  To relax
the LFV constraints one would need strong cancellations in the flavor violating entry, however
regions with large cancellations are not favored by a light Higgs mass of around 125 GeV. 
Partial cancellations are however allowed which put these regions within the reach of 
MEG (Project X) for $\mu \to e  \gamma$ ($\mu \to eee $).

\acknowledgments
AM would like to acknowledge support from the MIUR PRIN project  ``Matter-Antimatter Asymmetry, Dark Matter and Dark Energy in the LHC Era" and the EU project ``Unification in the LHC Era" contract PITN-GA-2009-237920 (UNILHC).
SKV thanks INFN, Padova section and Dipartimento di Fisica `Galileo Galilei', University of Padova, for supporting his visit. He also thanks DST Ramanujan Fellowship SR/S2/RJN-25/2008 of Government of India for support. LC and KMP are grateful to CHEP, IISc for hospitality and support during their visits. KMP also thanks Anjan S. Joshipura for partial support for this visit.  \appendix

\section{Description of Future Experiments and  Prospects: Circa 2020} %: Project X, PRISM/PRIME}
\label{sec6}

\begin{table}[t]
 \begin{center}
 \begin{tabular}{llcc}
 \hline
 \hline
 LFV process & Experiment & Future limits & Year (expected) \\
 \hline
 BR($\mu\rightarrow e \gamma$) 	& MEG~\cite{Adam:2011ch} & ${\cal O}(10^{-13})$ & $\sim$ 2013\\
				& Project X~\cite{projectx} & ${\cal O}(10^{-15})$ & $>$ 2021\\
 BR($\mu\rightarrow e e e$)   	& Mu3e~\cite{mu3e} & ${\cal O}(10^{-15})$ & $\sim$ 2017\\
				& ~~~'' & ${\cal O}(10^{-16})$ & $>$ 2017\\
				& MUSIC~\cite{Kuno:2011zz} & ${\cal O}(10^{-16})$ & $\sim$ 2017\\
				& Project X~\cite{projectx} & ${\cal O}(10^{-17})$ & $>$ 2021\\
 CR($\mu\rightarrow e$)        	& COMET~\cite{Kuno:2011zz} & ${\cal O}(10^{-17})$ & $\sim$ 2017\\
				& Mu2e~\cite{mu2e} & ${\cal O}(10^{-17})$ & $\sim$ 2020\\
				& PRISM/PRIME~\cite{Barlow:2011zz, Kuno:2011zz} & ${\cal
				  O}(10^{-18})$ & $\sim$ 2020\\
				& Project X~\cite{projectx} & ${\cal O}(10^{-19})$ & $>$ 2021\\
 BR($\tau\rightarrow \mu \gamma$) & Belle II \cite{Aushev:2010bq} & ${\cal O}(10^{-8})$ & $>$ 2020\\
 BR($\tau\rightarrow \mu \mu \mu$) & Belle II \cite{Aushev:2010bq} & ${\cal O}(10^{-10})$ & $>$ 2020 \\
 BR($\tau\rightarrow e \gamma$) & Super B~\cite{Hewett:2012ns} & ${\cal O}(10^{-9})$ & $>$ 2020\\
 BR($\tau\rightarrow \mu \gamma$) & Super B~\cite{Hewett:2012ns} & ${\cal O}(10^{-9})$ & $>$ 2020\\
 BR($\tau\rightarrow \mu \mu \mu$) & Super B~\cite{Hewett:2012ns} & ${\cal O}(10^{-10})$ & $>$ 2020\\
 \hline
 \hline
 \end{tabular}
 \end{center}
\caption{Future sensitivities of next-generation experiments.}
\label{nextgenexp}
\end{table}

PRISM \cite{Barlow:2011zz, Kuno:2011zz} or Phase Rotated Intense Slow Muon source is an upcoming
facility at J-PARC. It will accelerates muon beam using a magnetic field inside a muon storage ring
and will deliver $10^{12}$ $\mu/$second at an energy of few 10s of MeV. PRIME \cite{Barlow:2011zz,
Kuno:2011zz} is a $\mu-e$ conversion detector for PRISM designed to stop clumps of $\mu$ in a thin
foil, having an energy of about 20 MeV. Because of high monochromaticity, large pulse rate  and very
low background level it would reach a sensitivity of detecting branching ratios of ($\mu \rightarrow
e$) conversion in nuclei around $10^{-18}$ within few years of running. The PRIME will improve the
sensitivities to $\mu-e$ conversions by two orders of magnitude compared to the next-generation
experiment COMET (Coherent Muon to Electron Transition) experiment.
% Compared to PRIME, COMET uses a proton
% beam of energy 56 kW from the J-PARC main ring and it is aimed at a
% sensitivity of $10^{-16}$ in detecting ($\mu \rightarrow e$) conversion in
% nuclei.

Project X \cite{projectx}, proposed at Fermilab, is a next-generation experiment which has the
potential to deliver very high power unprecedentedly intense $\mu$-beams for precise measurements of
the rare muon decays. Project X will use 0.5 MW beam of $\mu$ accelerated at 3 GeV which will be
generated from high-power primary proton beam. With such an intense continuous muon beam,
Project X can look for $\mu\rightarrow e \gamma$ with sensitivity of ${\cal O}(10^{-15})$ which is
two order of magnitude improvement over the future reach of MEG. However, sensitivity below
$10^{-15}$ appears beyond the reach of Project X unless innovative ideas regarding the detectors
emerge. Like $\mu\rightarrow e \gamma$, searches for $\mu \rightarrow 3e$ also need continuous muon
beam. Experiments at Project X will improve the sensitivity to $\mu \rightarrow 3e$ decays by at
least three to four orders of magnitude, assuming significant efforts to develop new detector
technologies. The Project X proposals for studying the $\mu-e$ conversion in nuclei are particularly
interesting. Two distinct scenarios has been proposed. If the next-generation round of experiments
(COMET and Mu2e) observe any signal of $\mu-e$ conversion, then the available muon beams at
Project X would allow further precision studies with several $\mu-e$ events in different nuclei.
If no signal of $\mu-e$ will be found in next-generation experiments, Project X could reach the
sensitivities ${\cal O}(10^{-19})$ or beyond given some improvements in beam technologies. To
achieve these goals, a proposal to use muon storage ring installed in the muon beam line is under
consideration.

The Belle II \citep{Aushev:2010bq} is an upgrade of the existing Belle detector at the KEK B-factory in Japan. 
It will make use of the upgraded Super KEKB accelerator. 
The Belle II detector is expected to collect 40 times more luminosity ($8 \times 10^{35}\, {\rm cm}^{-2}\, {\rm s}^{-1}$) 
than the previous generation Belle detector and by  2022 it would collect an integrated luminosity of $\sim 50\, {\rm ab}^{-1}$. 
The sensitivity for the $\tau$ two body decays is expected to improved by an order of magnitude. 

Super B factory \cite{Hewett:2012ns} is a proposed high luminosity electron-positron collider. With
an integrated luminosity of 75 ab$^{-1}$, Super B would be able to explore a significant portion of
parameter space of new physics scenarios by searching for LFV in $\tau$ decays. It is going to
improve the sensitivities to different channels of $\tau$ decays by at least an order of magnitude
or two. The future sensitivities of all these experiments are summarized in the table \ref{nextgenexp}.

%====================================================================

\bibliographystyle{JHEP}

\bibliography{nuhm_draft}

\end{document}